\documentclass[11pt,a4paper]{article}

\usepackage{amssymb}
\usepackage{amsthm}
\usepackage[scaled=0.92]{helvet}
\usepackage{courier}
\usepackage{graphicx} 
\usepackage{amsmath}
\usepackage{bm}
\usepackage{amsfonts}
\usepackage{mathrsfs}
\usepackage{enumerate}
\usepackage{subfigure}
\usepackage{hyphenat} 
\usepackage{color}

\DeclareMathOperator*{\argmin}{arg\,min}

\begin{document}




\title{Comparison of several data-driven nonlinear system identification methods on a simplified glucoregulatory system example}


\author{Anna Marconato$^{a}$, Maarten Schoukens$^{a}$, Koen Tiels$^{a}$,\\ W. Dhammika Widanage$^{b}$, Amjad Abu-Rmileh$^{c}$, Johan Schoukens$^{a}$}

\date{}

\maketitle

{\center{ \small
$^{a}$ Dept. ELEC, Vrije Universiteit Brussel, Pleinlaan 2, 1050 Brussels, Belgium, \\ Email: anna.marconato@vub.ac.be, Fax: +32-2-629.28.50\\
$^{b}$ WMG, University of Warwick, Coventry, CV4 7AL, U.K.\\
$^{c}$ Department of Brain and Cognitive Sciences, \\ Ben-Gurion University of the Negev, Beer-Sheva, 84105 Israel
\\}}


\begin{abstract}

In this paper, several advanced data-driven nonlinear identification techniques are compared on a specific problem: a simplified glucoregulatory system modeling example. This problem represents a challenge in the development of an artificial pancreas for T1DM treatment, since for this application good nonlinear models are needed to design accurate closed-loop controllers to regulate the glucose level in the blood. Block-oriented as well as state-space models are used to describe both the dynamics and the nonlinear behavior of the insulin-glucose system, and the advantages and drawbacks of each method are pointed out. The obtained nonlinear models are accurate in simulating the patient's behavior, and some of them are also sufficiently simple to be considered in the implementation of a model-based controller to develop the artificial pancreas.

\end{abstract}




\section{Introduction}
\label{intro}

The availability of good nonlinear modeling techniques is highly beneficial in many branches of engineering. Nonlinear models are needed in various applications, e.g. to understand and analyze the system under test, to simulate the behavior of the device during the design phase, or to design and implement a controller in industrial processes.
System identification provides us with a variety of methods to derive accurate mathematical descriptions of the underlying system, based on a set of input/output measurements \cite{Lju10}.

Among the different choices for the (nonlinear) model structure, block-oriented models are simple representations that are easy to interpret, and for this reason they are often used in practice (see \cite{gir10} for an overview of recent advances). On the other hand, state-space models are very flexible descriptions, and offer the advantage of being more suited for Multiple-Input Multiple-Output (MIMO) problems \cite{ver02,pad10}. 

Given the particular problem under study, the user needs to select which model structure to adopt to describe the system's behavior. This choice is often made by trading-off model accuracy and complexity, taking into account the constraints of the problem. 

In this paper, several nonlinear block-oriented and state-space models are compared. Within the class of block-oriented models, single-branch and parallel Wiener models are considered. Within the state-space models, polynomial nonlinear state-space and neural network based state-space models are considered. All these nonlinear models are compared on a specific example, the identification of a simplified glucoregulatory system. This problem is a major challenge in the development of the artificial pancreas for Type 1 diabetes mellitus (T1DM) patients.

T1DM is a disease characterized by the fact that the pancreas is not able to produce a sufficient amount of insulin. When treating patients with exogenous insulin delivery, a closed-loop model-based controller serves as an essential component of the artificial pancreas that is responsible for regulating the level of glucose in the blood.
Accurate but simple models of the glucoregulatory system are needed to tune the artificial pancreas in real-time controllers.

Several mathematical descriptions (mainly first principle models) have been considered to represent a diabetic patient (e.g. the Meal model \cite{Dal07}, the Hovorka model \cite{hov04}, and the minimal modeling approach \cite{Berg79}). These models can be used to verify the quality of a controller, but they are not suited for online tuning. Examples of automated closed-loop control systems based on these models are currently under study. Extensive reviews of the different controller types used for the insulin-glucose problem, including PID controllers and Model Predictive Control (MPC) approaches, can be found in \cite{Cob09,Beq12}). Although the physical models retain physiological relevance and interpretability of the underlying system, they present the major limitation of being too complex for control and prediction algorithms design. 
Therefore, the application of advanced identification techniques to obtain simple behavioral models of insulin-glucose systems, based on available input/output data, represents a crucial step towards the development of the artificial pancreas for T1DM patients. 

In \cite{Amjad2012} it is shown that identification of nonlinear models is needed, since linear modeling does not guarantee an accurate approximation of the (nonlinear) glucoregulatory system, especially when the model is required for a wide operating range.

One way of extending the linear framework is to consider linear parameter varying (LPV) models, as done in \cite{cer12} where a set-membership LPV identification method is exploited to model the insulin-glucose dynamics.
A different approach, based on nonlinear modeling, is studied here. The reason why the nonlinear framework is chosen in this work is given by the intrinsic nonlinear nature of the first principle models (\cite{Dal07,hov04}). LPV models are typically suited to describe a nonlinear system only if the input signal stays rather small around a moving set point, since the nonlinear behavior can then be satisfactorily linearized with a good approximation. Their performance deteriorates if one tries to extend the use of LPV models to larger signals. In this paper, nonlinear models will be considered to capture as accurately as possible the nonlinear dynamics characterizing the physical descriptions. The considered nonlinear models are single-branch and parallel Wiener models, and polynomial and neural network based nonlinear state-space models.

The objective of this work is the comparison of several nonlinear identification techniques to obtain a (simple) behavioral model for the glucoregulatory system example. 

The focus is on the comparison of the different nonlinear identification techniques. Simulations of the Meal model \cite{Dal07} will be used as a benchmark for the different identification methods. Note that the glucoregulatory system is characterized by two inputs: the insulin delivered to the patient, and the meal intake. However, to simplify the analysis, in this study only the insulin will be considered as an input, since it is the main control variable. The meal intake will not be considered in this paper.

Block structures as well as nonlinear state-space representations are considered to model both the dynamics and the nonlinear behavior of the insulin-glucose system. To characterize the system nonlinearities, polynomials and nonlinear functions from statistical learning (namely neural networks) are used.

This paper is organized as follows. Section~\ref{sec2} introduces the considered problem and the objective of the work. Sections~\ref{sec3} and \ref{sec4} discuss the details of the block-oriented identification and nonlinear state-space modeling techniques that are then applied to the glucoregulatory system identification problem in Sections~\ref{sec5} and \ref{sec6}. A critical discussion of the results and of the computational complexity aspects, together with some issues about the controller implementation are presented in Section~\ref{sec7}. Finally, some conclusions are drawn in Section~\ref{concl}.

\section{Problem formulation}
\label{sec2}

In general terms, a system identification task consists in finding a mathematical description, i.e. a model, that characterizes the input/output relationship of the given system, on the basis of a set of measured values, and, if available, any \textit{a priori} information about the underlying system. Obtaining accurate models proves to be very useful for several purposes: to simulate the system behavior without the need of performing expensive experiments, to design or optimize industrial processes, to build a controller for a specific device, etc.

Many different techniques have been developed to solve system identification problems, and while the linear case is already well understood \cite{Lju99,PinSch12}, nonlinear modeling still offers interesting challenges and open problems (see \cite{Lju10} for a complete overview).

The application considered in this work is the identification of the glucoregulatory system. The idea is to take a first step towards the estimation of a model that can be used to implement a controller for the development of the artificial pancreas for T1DM patients. 

As mentioned above, the glucoregulatory system normally has two inputs: the insulin and the meal intake. Here a first step is taken in order to understand the behavior of one branch of the two-input system, namely the insulin-glucose subsystem. Note that, in this application, the insulin is the control variable, so it is natural to start addressing the problem by focusing on the insulin-glucose branch. 

Moreover, in \cite{Amjad2012}, it is shown that for a Meal model, used in this work to simulate the glucoregulatory system and to generate the benchmark data, the insulin-glucose subsystem exhibits a higher nonlinear behavior than the meal-glucose branch, which makes the identification of the insulin-glucose subsystem a more challenging problem.

Clearly, given the nonlinear nature of the glucoregulatory system, there might be mixing effects present between the two inputs. This would mean that one cannot simply decompose the two-input one-output system into two independent single-input single-output (SISO) parts.

However, in the field of closed-loop glucose control, it is a common practice to obtain a SISO insulin-glucose model for the design of feedback control algorithms (see e.g. \cite{March08} and \cite{vHeus12}). 
If a feedfoward controller is considered, then another SISO meal-glucose model (representing the disturbance effect) is also identified. Considering the nonlinearity in the insulin-glucose model proved to enhance the controller performance, while the insertion of nonlinearity in formulating a feedforward control part would increase the complexity, without significant improvement in the overall performance \cite{AmjadGa12}.

Furthermore, studying one branch is useful to get insight about the partial behavior of the system. This choice provides us with a good benchmark, linked to a relevant application, on which several nonlinear models can be compared.


Therefore, in this paper the focus will only be on the simpler SISO case.

\begin{figure}[!t]
\centering
\includegraphics[width=7cm]{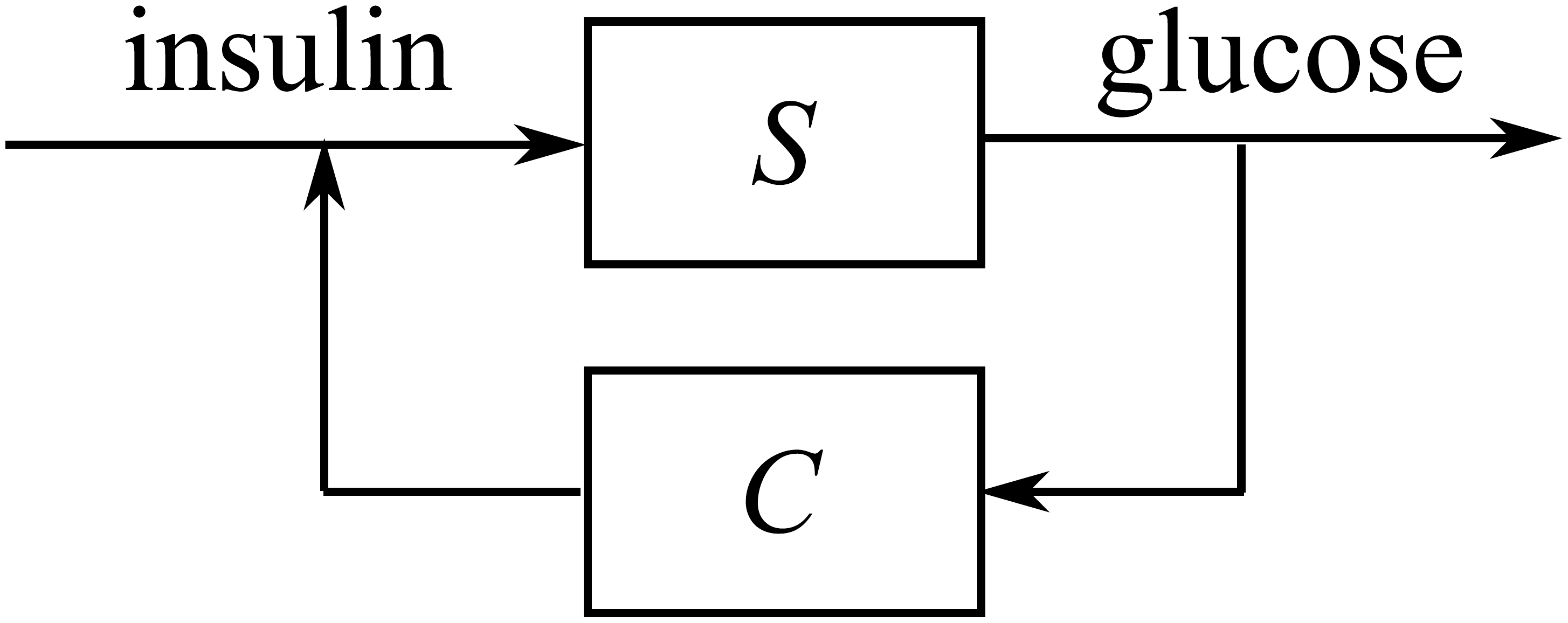}
\caption{Schematic representation of the considered simplified glucoregulatory system (S). A model-based controller (C) is needed to regulate the level of glucose in the patient's blood.}
\label{figsyscontr}
\end{figure}

Figure~\ref{figsyscontr} depicts the considered simplified glucoregulatory system and the controller. 
In this particular example, the input to the system is the insulin delivered to the patient, and the output is the glucose level. The aim of the controller is then to automatically compute the required insulin dose to keep the concentration of glucose in the blood within the desired range. 

To be used for control purposes, the identified model should be: (1) as accurate as possible, since large errors may cause serious damage to the patient; (2) not too complex, to be able to easily implement the controller.

A first step in the identification procedure can be to build a good linear model and check whether this description is sufficient to characterize the system. Linear models are indeed easy to interpret, and quite simple to  obtain and to use. Moreover, several well-established techniques exist for linear identification, both in the time- \cite{Lju99} and in the frequency domain \cite{PinSch12}. 

In this work, the Best Linear Approximation (BLA) -- introduced in more details in the next subsection -- is estimated and used as starting point for all the nonlinear identification methods that will be presented.

\subsection{Best Linear Approximation (BLA)}

In the class $\mathscr{G}$ of all linear models, the BLA is defined to be optimal in least square sense:
$$\hat{G}_{BLA}=\argmin_{G\in\mathscr{G}}{E\{\left|y(t)-G(u(t))\right|^2\}}$$
Here $u(t)$ and $y(t)$ are the input and output of the considered system \cite{PinSch12}. 

As a result, a nonlinear system can be described as a linear system plus a noise source representing all nonlinear distortions and noise contributions that are not captured by the BLA.


To obtain a good linear description of the underlying nonlinear system, first the Frequency Response Function (FRF) $\hat{G}_{BLA}(j\omega_k)$ at frequencies $\omega_k$ is estimated nonparametrically. This can be done in the frequency domain by applying the Local Polynomial Method \cite{Pin10a,PinSch12}, which allows one to remove transient terms and to get an estimate of the sample variance $\hat{\sigma}_{\hat{G}_{BLA}}^2(j\omega_k)$ of the nonparametric BLA.

As a next step, a parametric transfer function $\hat{G}_{BLA}(j\omega_k,\boldsymbol{\theta})$ is obtained starting from the estimated FRF, e.g. using the ELiS toolbox \cite{elis}:
\begin{equation}
\label{eqBLApar}
\hat{G}_{BLA}(j\omega_k,\boldsymbol{\theta})=\frac{b_0+b_1q^{-1}+\cdots+b_{n_b}q^{-n_b}}{a_0+a_1q^{-1}+\cdots+a_{n_a}q^{-n_a}}
\end{equation}
where $q^{-1}$ is the backward shift operator, and the parameter vector $\boldsymbol{\theta}$ contains all $a_i$ and $b_j$ values.

The sample variance $\hat{\sigma}_{\hat{G}_{BLA}}^2(j\omega_k)$ estimated in the previous step can be used as a weight term to define a weighted cost function in the optimization routine as follows:

$$V(\boldsymbol{\theta})=\sum_{k=1}^{F}{\frac{\left|\hat{G}_{BLA}(j\omega_k)-\hat{G}_{BLA}(j\omega_k,\boldsymbol{\theta})\right|^2}{\hat{\sigma}_{\hat{G}_{BLA}}^2(j\omega_k)}}$$

\subsection{Linear vs. nonlinear modeling}

\begin{figure}[!t]
\centering
\includegraphics[width=11cm]{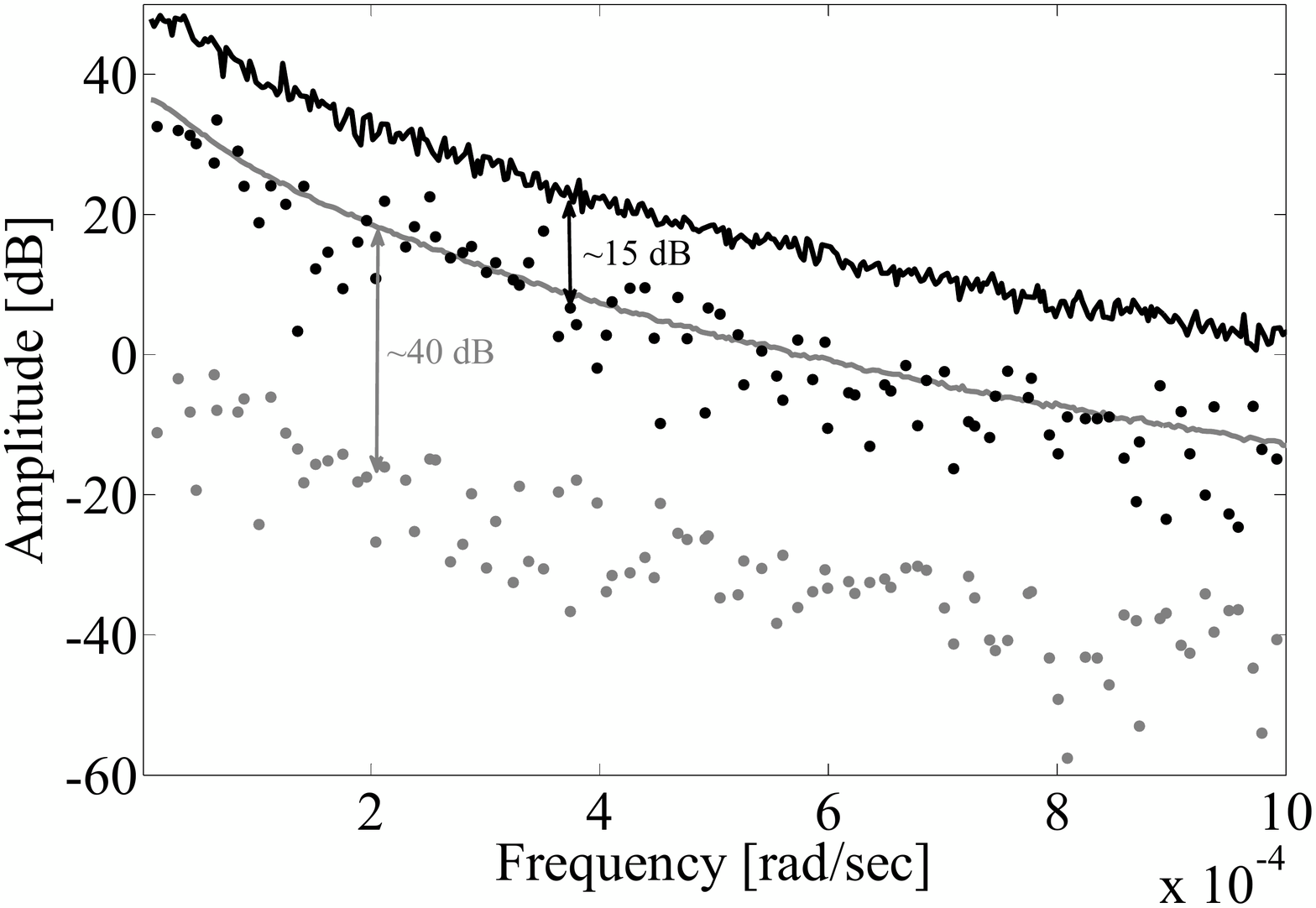}
\caption{Nonlinearity level for an example of an insulin-glucose system. The output spectrum (solid line) and the nonlinear distortions (bullets) are shown, for two different excitation levels (black: high level, gray: low level) \cite{Amjad2012}.}
\label{figNLlevel}
\end{figure}

Figure~\ref{figNLlevel} shows the nonlinearity level of a typical example of an insulin-glucose system \cite{Amjad2012}. 
On this occasion the input is a signal known as a random phase multisine \cite{PinSch12}. It is a periodic broadband signal designed with uniform power over the bandwidth of interest but has several harmonics suppressed to detect the level of nonlinearity \cite{Evans94}. More in details, a random phase multisine with period length $N$ is defined as follows:
\begin{equation}
\label{eqRPM}
u_{RPM}(t)=\sum_{k=1}^{F}{A_k\cos{(2\pi k \frac{t}{N}+\varphi_k)}}
\end{equation}
where $F$ is the number of excited frequencies, and the phases $\varphi_k$ are random variables that are independent over the frequency and uniformly distributed in the interval $[0,2\pi)$.
Two rms levels of a random phase multisine excitation are considered to analyze the nonlinearity level of the system. For a low input rms level the nonlinear contributions are about 30-40 dB lower than the output spectrum (i.e. a linear model can give a satisfactory description of the system behavior), but when a higher input rms level is considered the nonlinearity level increases significantly. In practice, this means that relying on a linear model would result in large error contributions that will affect the performance of the controller, potentially exposing the patient to high risks.

This motivates the application of advanced nonlinear modeling techniques to solve the simplified SISO glucoregulatory system identification problem. 

In this paper two main classes of approaches are discussed: block-oriented modeling and nonlinear state-space modeling. Some of the most recent advances within these two groups of methods will be presented in more details in the next two sections, and will then be applied to the considered insulin-glucose problem.

\section{Block-oriented identification}
\label{sec3}

In the recent years, there has been an increasing interest in block-oriented nonlinear modeling \cite{gir10}. Several classes of block-oriented models are obtained by combining two basic block types in different ways: linear time-invariant (LTI) blocks and static nonlinear blocks. 

Due to the simplicity of block structures and the fact that they are easy to interpret, block-oriented models are often used in practice. Many identification methods have been proposed that address a specific model class (Wiener, Hammerstein, Wiener-Hammerstein, and so on, see works in \cite{gir10} for some examples). In the following, two types of block-structured models are discussed in more details: Wiener models and Wiener-Schetzen models.

\subsection{Wiener models}
\label{secWiener}

\begin{figure}[!t]
\centering
\subfigure[]{
\includegraphics[width=8cm]{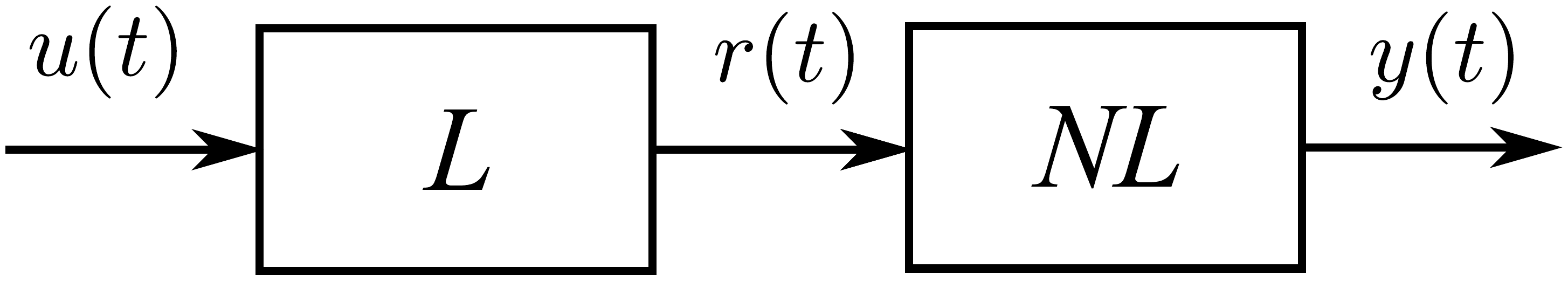}
\label{figW}
}
\subfigure[]{
\includegraphics[width=8cm]{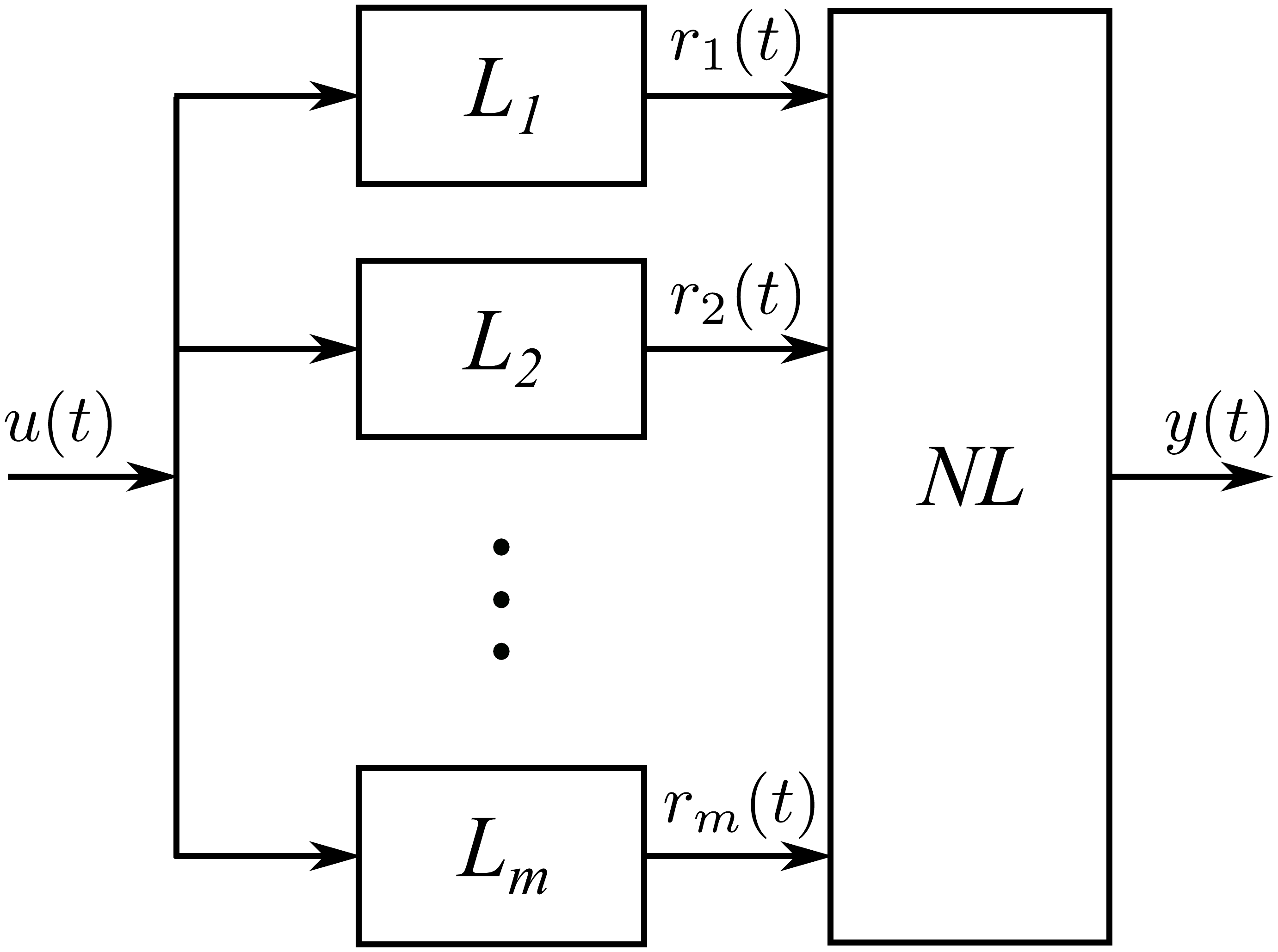}
\label{figWpar}
} 
\caption{(a) Wiener models. (b) Parallel Wiener models.}
\end{figure}

By connecting a linear block in series with a static nonlinearity one obtains the class of Wiener models (Figure~\ref{figW}), which has been successfully applied to describe the input/output relationship of many nonlinear systems, see e.g. \cite{Hun86,Zhu99}. 


The identification algorithm for the Wiener structure consists of three steps \cite{Cra01}:

\begin{enumerate}

\item Identification of the linear dynamic block by estimating the BLA. In practice this is done by applying first the Local Polynomial Method \cite{Pin10a,PinSch12} and then estimating a parametric linear model (as in Eq.~(\ref{eqBLApar})) starting from the nonparametric BLA. When a Gaussian input is used, the dynamics estimated with the BLA are equal to the LTI block of a Wiener system \cite{PinSch12}. In particular, the BLA of a Wiener system is given by:
\begin{equation}
\label{eqBLAWiener}
\hat{G}_{BLA}(q)=\alpha L(q)
\end{equation}
where the real coefficient $\alpha$ depends on the system and on the power spectrum of the Gaussian input signal, and $L(q)$ is the LTI block.

\item Identification of the static nonlinear block. By filtering the input signal with the estimate of the linear block obtained in the previous step, the input $r(t)$ to the nonlinear block is known up to the scale factor $\alpha$. Note that knowledge of this scale factor is not required, as an arbitrary nonzero gain can be exchanged between the linear and nonlinear block without affecting the input/output behavior of the system. Hence the static nonlinearity can be easily modeled. Here two different choices for the nonlinearity are considered:

\begin{itemize}

\item a polynomial representation:
\begin{equation}
\label{eqpoly}
y(t)=\sum_{k=0}^{d}{\beta_k r^k(t)}
\end{equation}
Note that this formulation is linear in the parameters $\beta_k$. This makes it possible to initialize the nonlinear term by solving a least squares problem in a fast and efficient way.

\item a one-hidden-layer feedforward sigmoidal neural network \cite{has09}:
\begin{equation}
\label{eqsigNN}
y(t)=\sum_{k=1}^{n}{\gamma_k \tanh(v_k r(t)+w_k)}+c
\end{equation}
Here the parameters $\gamma_k,v_k,w_k,c$ appear nonlinearly, therefore a nonlinear optimization routine is needed to initialize a nonlinear block with this structure.

\end{itemize}

\item Optimization of all parameters via a nonlinear least squares optimization routine. This step is computationally the most expensive part of the identification algorithm for Wiener models.

\end{enumerate}


\subsection{Wiener-Schetzen models}
\label{secWienerSch}

To increase the model flexibility, one can also decide to add a number of different Wiener models in parallel, obtaining the class of parallel Wiener models (Figure~\ref{figWpar}). A subset of the parallel Wiener class is given by Wiener-Schetzen models. Here, the linear dynamic blocks are rational orthonormal basis functions \cite{Heu05}, and the nonlinearity is a multivariate polynomial. In spite of its simple structure, this model class has proven to approximate all fading memory open loop nonlinear systems with an arbitrarily high accuracy \cite{Boy85}.


In this work, the Wiener-Schetzen model is estimated as follows \cite{Koen11}:

\begin{enumerate}

\item Determination of the orthonormal basis functions $L_i, i=1,\ldots,m$, based on the poles of the parametric BLA. This is done, for each branch, as follows:
\begin{equation}
\label{eqorthbf}
L_i(q)=\frac{\sqrt{1-\left|p_i\right|^2}}{q-p_i}\prod_{j=1}^{i-1}{\frac{1-p_j^*q}{q-p_j}}
\end{equation}
where the $p_i$'s are the poles of the parametric BLA. 

\item Identification of the multivariate polynomial function (after filtering the input signal $u(t)$ through the different estimated linear blocks, to obtain the signals $r_i(t), i=1,\ldots,m$):
\begin{equation}
\label{eqmultipoly}
y(t)=\sum_{s=0}^{d}{\sum_{k_1+\cdots+k_m=s}{\beta_{k_1,\ldots,k_m}\prod_{i=1}^{m}{r_i^{k_i}(t)}}}
\end{equation}
The parameters $\beta_{k_1,\ldots,k_m}$ of the nonlinear part are obtained by minimizing the mean square error between the modeled output and the true output $y(t)$, which is a problem that is linear in the parameters, and that therefore can be solved easily and fast.

\end{enumerate}

When no noise is present, the root mean square error on the modeled output can be made arbitrary small by increasing the number of basis functions in the linear part, at the cost of a significant increase in the number of parameters of the model. In this way, there is no need to further optimize the parameters with a nonlinear optimization routine. Note, however, that the computational complexity increases combinatorially with the number of branches $m$, since the nonlinear term is formulated as a multivariate polynomial.

\section{Nonlinear state-space models}
\label{sec4}

A different modeling approach to characterize the behavior of nonlinear dynamic systems is the identification of nonlinear state-space models. These are black-box models that are described by two equations: a state equation (\ref{eq1a}), and an output equation (\ref{eq1b}):

\begin{align}
\mathbf{x}(t+1)&=f(\mathbf{x}(t),\mathbf{u}(t)) \label{eq1a}\\
\mathbf{y}(t)&=g(\mathbf{x}(t),\mathbf{u}(t)) \label{eq1b}
\end{align}

where $\mathbf{u}(t)$ and $\mathbf{y}(t)$ are the input and output signal respectively, $\mathbf{x}(t)$ is the state sequence that represents the memory of the system and is in general unknown, and $f(\cdot)$ and $g(\cdot)$ are the nonlinear functions to be estimated. 

Nonlinear state-space models are flexible model structures that represent naturally system dynamics, they allow one to easily deal with multiple-input multiple-output (MIMO) systems and have already been used to successfully describe a variety of systems \cite{Pad08,annaCEPWH}. In contrast to the previous model structures, nonlinear state-space models can also describe systems with nonlinear feedback.

In this paper, two approaches are discussed, which differ in the choices that are made to describe the nonlinear parts in the model, and in the identification techniques that are developed to solve the modeling task: a polynomial state-space model and a neural network-based state-space model.

\subsection{Polynomial Nonlinear State-Space (PNLSS) models}
\label{secPNLSS}

PNLSS models extend the classical linear state-space representation by adding polynomial terms, as in the following equations:

\begin{align}
\mathbf{x}(t+1)&=\mathbf{A}\mathbf{x}(t)+\mathbf{B}\mathbf{u}(t)+\mathbf{E}\boldsymbol{\zeta}(\mathbf{x}(t),\mathbf{u}(t)) \label{pss1}\\
\mathbf{y}(t)&=\mathbf{C}\mathbf{x}(t)+\mathbf{D}\mathbf{u}(t)+\mathbf{F}\boldsymbol{\eta}(\mathbf{x}(t),\mathbf{u}(t)) \label{pss2}
\end{align}
where $\mathbf{A}$, $\mathbf{B}$, $\mathbf{C}$ and $\mathbf{D}$ are the coefficient matrices characterizing the linear part of the model, vectors $\boldsymbol{\zeta}$ and $\boldsymbol{\eta}$ contain monomials in $\mathbf{x}(t)$ and $\mathbf{u}(t)$ (up to a user-specified degree), and matrices $\mathbf{E}$ and $\mathbf{F}$ contain the coefficients corresponding to these monomials.

The identification procedure for PNLSS models consists of the following steps \cite{pad10}:

\begin{enumerate}

\item Nonparametric estimation of the BLA.

\item Transformation of the nonparametric estimate into a parametric linear model by means of the frequency domain subspace algorithm \cite{Mkel96}. In this way, an initial estimate of matrices $\mathbf{A}$, $\mathbf{B}$, $\mathbf{C}$ and $\mathbf{D}$ is obtained.

\item Nonlinear least squares optimization of the linear model parameters using the Levenberg-Marquardt algorithm \cite{PinSch12}, with the estimates from the previous step as initial guesses for the optimization. 

\item Estimation of the nonlinear model. The Levenberg-Marquardt algorithm is used to minimize a weighted least squares cost function with respect to all model parameters, i.e. matrices $\mathbf{A}$, $\mathbf{B}$, $\mathbf{C}$ and $\mathbf{D}$ for the linear part, and matrices $\mathbf{E}$ and $\mathbf{F}$ for the nonlinear part. The inverted noise covariance matrix is typically used as weighting factor. This optimization step is computationally very expensive. In particular, what slows down significantly the computation is the recursive nature of the Jacobian evaluation at each iteration of the optimization algorithm. Computing the Jacobian (with respect to the parameters appearing in the state equation (\ref{pss2})) is equivalent to calculating (in a recursive way) the output of an alternative PNLSS model. This impacts strongly on the total complexity of the identification algorithm.

\end{enumerate}

PNLSS models have shown to be a very flexible structure capable of describing a large number of nonlinear systems. However, when increasing the nonlinear degree, the number of parameters grows combinatorially, hence in practice the nonlinear degree should be chosen to be relatively low.

\subsection{Neural network-based state-space models}
\label{secNN-NLSS}

An alternative to model nonlinear systems within the state-space framework is to consider other kinds of nonlinear functions, for example nonlinearities coming from the statistical learning community, such as neural networks \cite{has09}, support vector machines \cite{vap98,suy02}, etc.

Here the following formulation of the nonlinear state-space problem is considered:

\begin{eqnarray}
\mathbf{x}(t+1)&=&f(\mathbf{x}(t),\mathbf{u}(t))= \nonumber \\ &=&\mathbf{A}\mathbf{x}(t)+\mathbf{B}\mathbf{u}(t)+f_{NL}(\mathbf{x}(t),\mathbf{u}(t)) \label{sstot1}\\
\mathbf{y}(t)&=&g(\mathbf{x}(t),\mathbf{u}(t))= \nonumber \\ &=&\mathbf{C}\mathbf{x}(t)+\mathbf{D}\mathbf{u}(t)+g_{NL}(\mathbf{x}(t),\mathbf{u}(t)) \label{sstot2}
\end{eqnarray}

where matrices $\mathbf{A}$, $\mathbf{B}$, $\mathbf{C}$ and $\mathbf{D}$ describe the linear part of the model and $f_{NL}$ and $g_{NL}$ are the nonlinear terms. Notice that, as for the PNLSS case in Eqs.(\ref{pss1}-\ref{pss2}), the separation between linear and nonlinear terms allows them to be identified in two steps.

The approach discussed here is based on the idea of cutting the recursion in the state equation (\ref{sstot1}), to obtain an approximate static version of the general nonlinear dynamic problem \cite{annaCEPWH}. Static regression methods can then be easily applied to model separately the nonlinear terms $f_{NL}$ and $g_{NL}$, making the whole estimation procedure faster and more efficient.

More in details, the identification algorithm is as follows:

\begin{enumerate}

\item Estimation of the BLA to obtain an initial estimate of matrices $\mathbf{A}$, $\mathbf{B}$, $\mathbf{C}$ and $\mathbf{D}$.

\item  Approximation of the unknown nonlinear state $\mathbf{x}(t)$ by solving a least squares problem, expressed as the trade-off between data fit and linear model fit, as follows \cite{annaCEPWH}:
\begin{align}
\hat{\mathbf{x}}(t)=& \arg\min_{\{ \mathbf{x}(t)\}} \sum_t{(\mathbf{y}(t)-\hat{\mathbf{C}}\mathbf{x}(t)-\hat{\mathbf{D}}\mathbf{u}(t))^2}  \nonumber \\ & + \lambda \sum_t{(\mathbf{x}(t+1)-\hat{\mathbf{A}}\mathbf{x}(t)-\hat{\mathbf{B}}\mathbf{u}(t))^2} \nonumber
\end{align}
where the matrices $\hat{\mathbf{A}}$, $\hat{\mathbf{B}}$, $\hat{\mathbf{C}}$ and $\hat{\mathbf{D}}$ are the ones estimated in the previous step, and $\lambda$ is a trade-off parameter that regulates the importance of the two criteria. This step is computed efficiently, thanks to the sparsity of the considered least squares problem.

Once an estimate $\hat{\mathbf{x}}(t)$ is obtained, it is possible to transform equation (\ref{sstot1}) into a static problem, by simply substituting the obtained values $\hat{\mathbf{x}}(t)$ at different time instants.

\item Estimation of the nonlinear terms $f_{NL}$ and $g_{NL}$ as two independent one-hidden-layer feedforward sigmoidal neural networks (expressed in a similar form as Eq.(\ref{eqsigNN})). In this way one obtains good initial values for the model parameters. This step requires only a static estimation of the nonlinear terms. However, the problem is nonlinear in the parameters, and therefore a nonlinear optimization routine is needed to complete the initialization.

\item Optimization of all model parameters using the Levenberg-Marquardt algorithm, after having reintroduced the dynamics in the general nonlinear state-space problem (\ref{sstot1}-\ref{sstot2}). This results again in a computationally very expensive step (as in the PNLSS case), due to the recursiveness in the Jacobian calculation.

\end{enumerate}

Notice that in practice, by tuning the number of sigmoid functions in the neural networks, it is possible to obtain an accurate model characterized by a reduced number of parameters.

\section{The insulin-glucose modeling problem}
\label{sec5}

In this and in the next section, the nonlinear identification methods described above are applied to the problem of modeling the insulin-glucose behavior for T1DM patients.

The Meal model \cite{Dal07} is employed here as a simulation model to generate the data. To produce the input/output data for the insulin-glucose subsystem, i.e. the branch of the glucoregulatory system on which we focus in this study, the meal intake (the second input signal in the Meal model) is set here equal to zero.

Note that the physical model is only used to reproduce the patient's input/output relationship for data generation purposes, while in the following the simpler behavioral models presented in Sections~\ref{sec3} and \ref{sec4} are used to estimate the system input/output behavior, based on the generated data.

\begin{figure}[!t]
\centering
\subfigure[]{
\includegraphics[width=8cm]{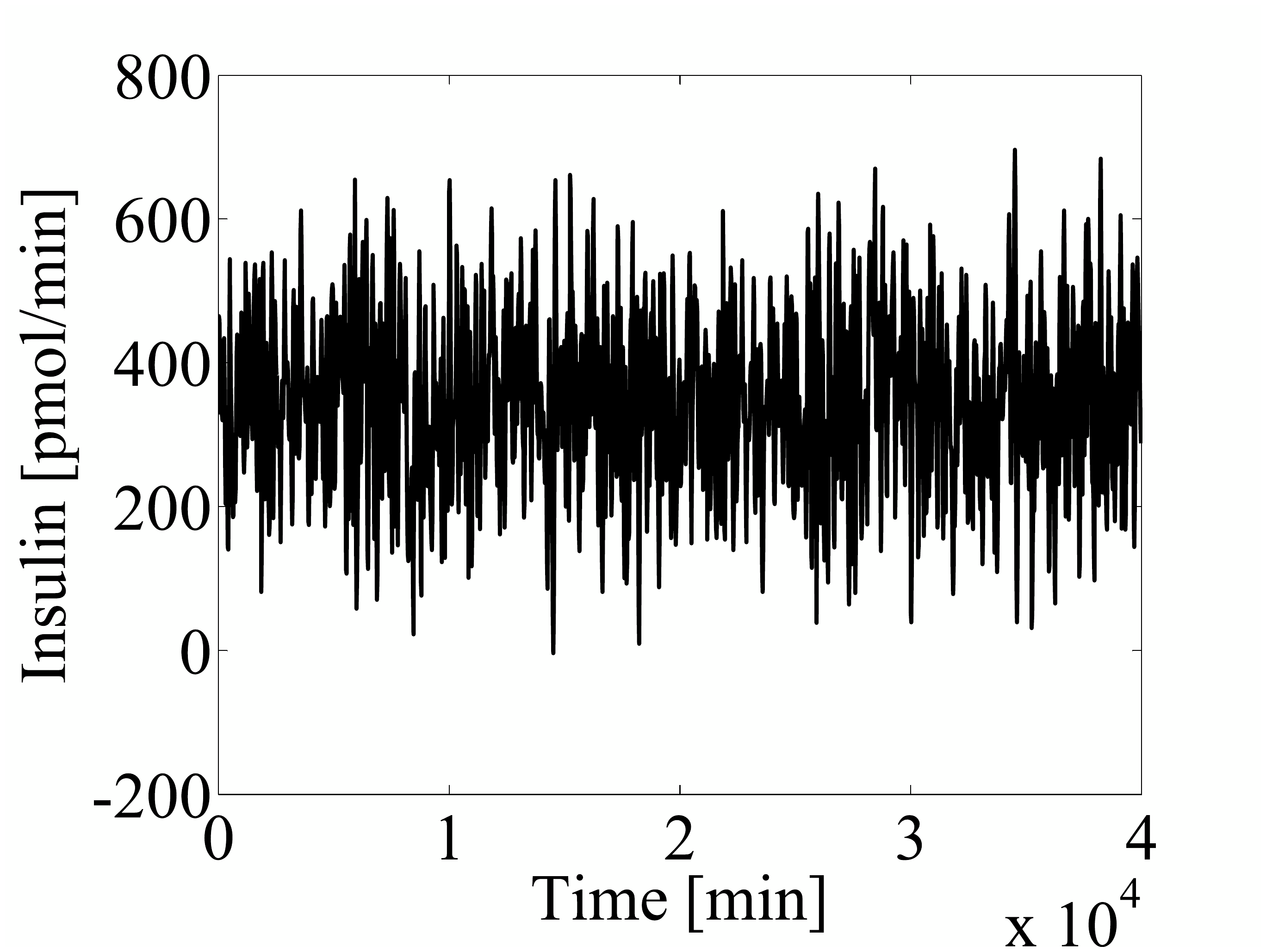}
\label{figexc}
}
\subfigure[]{
\includegraphics[width=8cm]{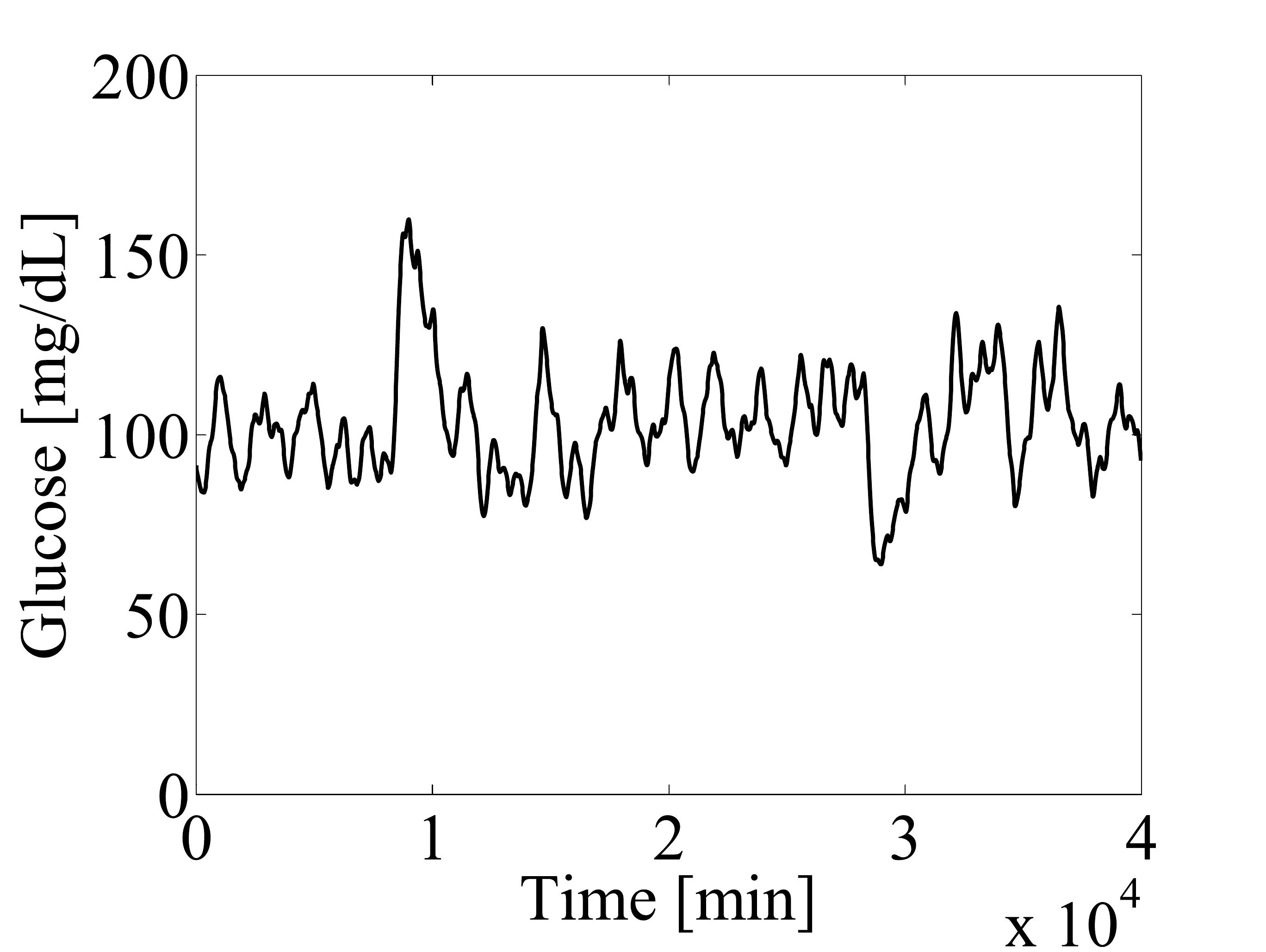}
\label{figout}
} 
\caption{(a) Excitation signal: random phase multisine (one period) at one operating point of the system. (b) Output signal, obtained by simulating the Meal model with the excitation signal above as input.}
\end{figure}

In this first example, the input signal $u(t)$ in Figure~\ref{figexc} is used to excite the system. It is a random phase multisine as in Eq.~(\ref{eqRPM}), with a sampling time of 20 minutes and a period length of 2000 points. The corresponding output signal $y(t)$ is shown in Figure~\ref{figout}. 

The random phase multisine excitation signal is chosen since it helps to reveal and study the dynamics of the system, while a more realistic type of signal will be considered later on in Section~\ref{secrealistic}. 

The multisine input signal is applied at 12 different operating points of the system, ranging from 100 to 550 pmol/min, i.e. 12 different offset levels of the input (i.e. basal insulin) are considered. At every operating point two periods of the signal are available. Two datasets are generated taking into account two different multisine rms levels: the first one used as estimation set, to build the models, and the second as validation set, to test the performance of the models on previously unseen data. 

Note that the large time scale of the signals considered in this example is used only for the simulation and study of the different nonlinear models.

%


\section{Results}
\label{sec6}

The results of the different methods are expressed in terms of the following relative error criterion, calculated at each of the 12 operating points:
$$e_{\text{rel}}(\%)=\frac{\left\|y(t)-\hat{y}(t)\right\|}{\left\|y(t)-\overline{y}\right\|}\cdot 100$$

where $\hat{y}(t)$ is the modeled output, $\overline{y}$ is the mean of the output over time at each operating point, and $\left\|\cdot\right\|$ is the Euclidean norm.

The number of model parameters is also important for the specific application, since it gives an indication of the model complexity, which should be kept low for control purposes.

\subsection{Best Linear Approximation}

\begin{figure}[!t]
\centering
\includegraphics[width=8cm]{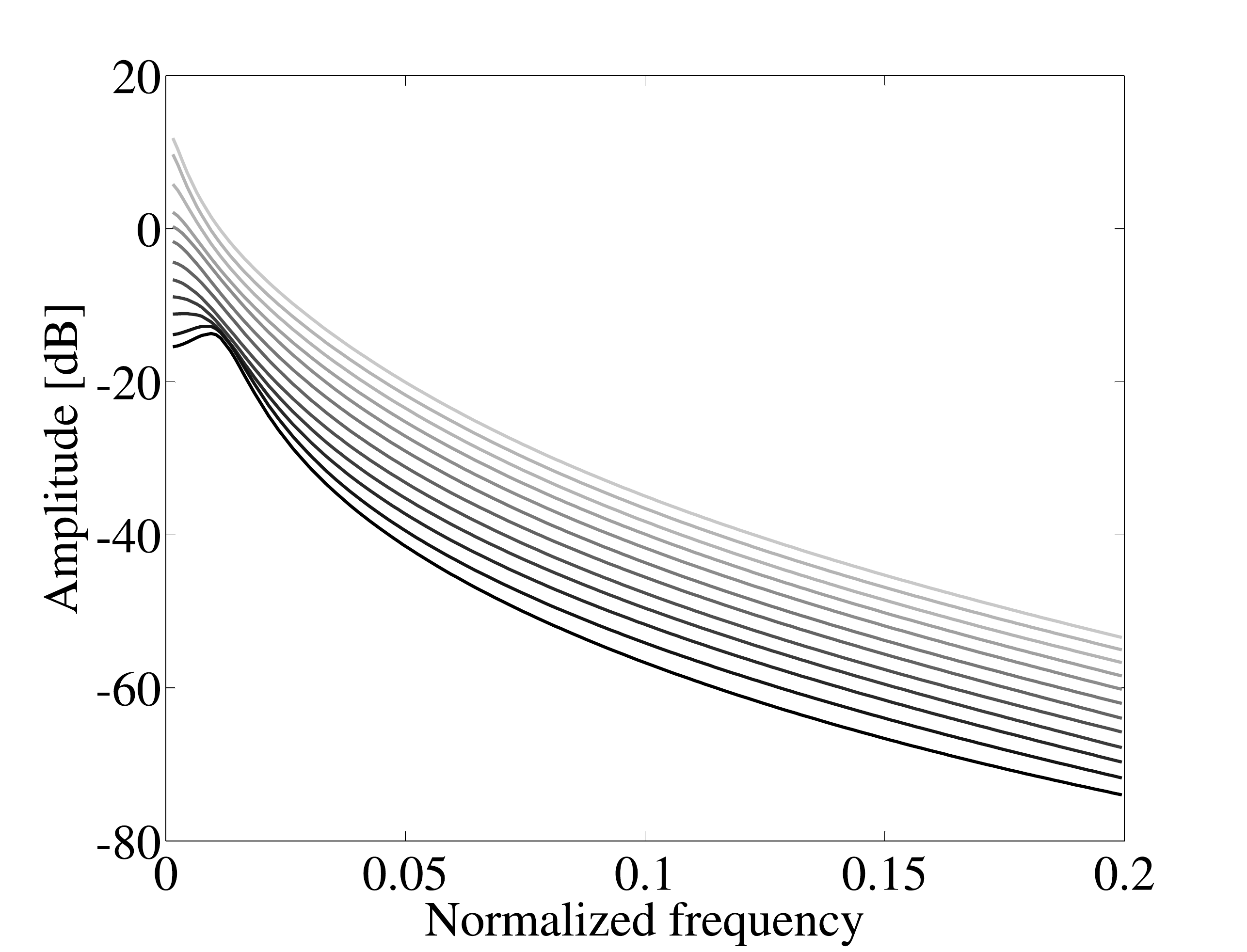}
\caption{Magnitude response of the transfer function for 12 operating points. Light gray to black line for operating points from 1 to 12.}
\label{figbla12}
\end{figure}

The magnitude response of a third order transfer function ($n_a=n_b=3$ in Eq.~(\ref{eqBLApar})) for the 12 operating points is shown in Figure~\ref{figbla12}.

It can be observed that there is some change in the amplitude and shape of the linear models at different operating points, which is due to the nonlinear behavior of the system.

To illustrate the performance of the linear model, the BLA computed at the 7th operating point is used to simulate the validation data across all the operating points. The corresponding relative error of the BLA is reported in Table~\ref{tabbla}.

\begin{table}[!t]
\begin{center}
\caption{\label{tabbla} {Results of the BLA computed at operating point 7 on the validation set, for the different operating points (o/p).}}
\begin{tabular}{ c  c  c  c  c  c  c  c  c  c  c  c  c }
\hline
\scriptsize{\textbf{o/p}} & \scriptsize{1} & \scriptsize{2} & \scriptsize{3} & \scriptsize{4} & \scriptsize{5} & \scriptsize{6} & \scriptsize{7} & \scriptsize{8} & \scriptsize{9} & \scriptsize{10} & \scriptsize{11} & \scriptsize{12} \\
\hline
\scriptsize{\textbf{$e_{\text{rel}}$} ($\%$)} & \scriptsize{78.3} & \scriptsize{70.4} & \scriptsize{60.0} & \scriptsize{47.1} & \scriptsize{37.3} & \scriptsize{19.4} & \scriptsize{7.0} & \scriptsize{27.5} & \scriptsize{52.8} & \scriptsize{80.0} & \scriptsize{108.2} & \scriptsize{137.0} \\
\hline
\end{tabular}
\end{center}
\end{table}


The BLA performs very good only around operating point 7, and it fails completely in the regions further away from it. Note that the relative error values larger than 100$\%$ at operating points 11 and 12 are due to the fact that the linear model error is larger than the signal.

Similar conclusions can be drawn when the BLA at another operating point is chosen. Therefore, it can be stated that a single LTI model is not suited to describe the overall input/output behavior of the system.

The obtained BLA is then used as a first step in the identification methods presented in Sections~\ref{sec3} and \ref{sec4} to build nonlinear models that describe the glucoregulatory system in a more accurate way, at all operating points.

\subsection{Nonlinear models}

Table~\ref{tabmodels} summarizes the main features of the nonlinear models obtained by applying the different methods on the insulin-glucose modeling problem (see Section number in Table~\ref{tabmodels} for the details).

\begin{table}[!t]
\begin{center}
\caption{\label{tabmodels} {Characteristics of the nonlinear models. Two types of nonlinear functions are used: polynomials (poly) and $\tanh(\cdot)$ sigmoid functions (tanh). $d$ represents the degree of the polynomials, while $n$ is the number of neurons in the hidden layer of the NN. For the NN-based NLSS, $n_{f}$ and $n_{g}$ are the number of neurons used in the state and output equation respectively. Remark: the linear part in the Wiener-Schetzen model consists of 10 parallel branches.}}
\begin{tabular}{ c  c  c | c }
\hline
\textbf{Model} & \textbf{BLA order} & \textbf{Nonlinearity} & \textbf{Section}\\
\hline
Wiener & 4 & poly, $d=4$ & \ref{secWiener} \\
\hline
Wiener-NN & 3 & tanh, $n=4$ & \ref{secWiener} \\
\hline
Wiener-Schetzen & 4 & poly, $d=2$ & \ref{secWienerSch} \\
\hline
PNLSS & 3 & poly, $d=3$ & \ref{secPNLSS} \\
\hline
NN-NLSS & 3 & tanh, $n_{f}=3$ $n_{g}=4$ & \ref{secNN-NLSS} \\
\hline
\end{tabular}
\end{center}
\end{table}






The obtained results are shown in Table~\ref{tabres}. The performance of the different nonlinear models is compared in terms of relative error on the validation set and number of parameters. The BLA results are also reported, as a reference.

\begin{table}[!t]
\begin{center}
\caption{\label{tabres} {Results of the nonlinear models: multisine excitation. The average relative error over the 12 operating points on the validation set is reported, together with the minimum and maximum relative error values (in brackets) and the number of model parameters.  As a comparison, also the results of a third order BLA are shown.}}
\begin{tabular}{ c  c  c }
\hline
\textbf{Model} & \textbf{Average $e_{\text{rel}}$ (min -- max) $\%$} & \textbf{$\#$ parameters} \\
\hline
Wiener & 21.1 (8.2 -- 33.8) & 15 \\
\hline
Wiener-NN & 14.6 (3.8 -- 37.0) & 21 \\
\hline
Wiener-Schetzen & 25.2 (2.2 -- 46.1) & 111 \\
\hline
PNLSS & 19.4 (8.9 -- 44.9) & 48 \\
\hline
NN-NLSS & 17.2 (7.2 -- 30.8) & 68 \\
\hline
\hline
BLA & 60.4 (7.0 -- 137.0) & 8 \\
\hline
\end{tabular}
\end{center}
\end{table}

A first observation that can be made is that the proposed nonlinear identification methods allow one to significantly improve the results obtained with the BLA. In particular, the average relative error over the 12 operating points is reduced approximately by a factor 3 for all nonlinear models.
Moreover, with the obtained nonlinear models it is possible to get low error results at all operating points, which was not the case for the linear model. 

A note on the number of parameters for the Wiener-Schetzen model: the large value is due to the fact that the output of the obtained model actually consists of a weighted sum of the outputs of three different nonlinear submodels. This was necessary in order to build a single nonlinear model characterized by good performance at all operating points.

Finally, a Wiener model can be as well described with a general NLSS representation, as expressed in Eqs.(\ref{eq1a}-\ref{eq1b}). Following this theoretical reasoning, it may appear surprising to note that the results obtained with the NN-NLSS approach are worse than for the Wiener-NN model. However, it is worth remarking that the higher error values for the NN-NLSS model are likely due to the presence of local minima issues during the nonlinear optimization step. 

This shows that it is important for the user to select a good model structure. A good model structure is not necessarily a very general one, but rather one that only has the complexity needed to describe the system under study. The more local minima are introduced by a (more complex) model structure, the higher the quality of the initial estimates needs to be to yield a satisfactory model.

\subsection{A more realistic case}
\label{secrealistic}

\begin{figure}[!t]
\centering
\includegraphics[width=8cm]{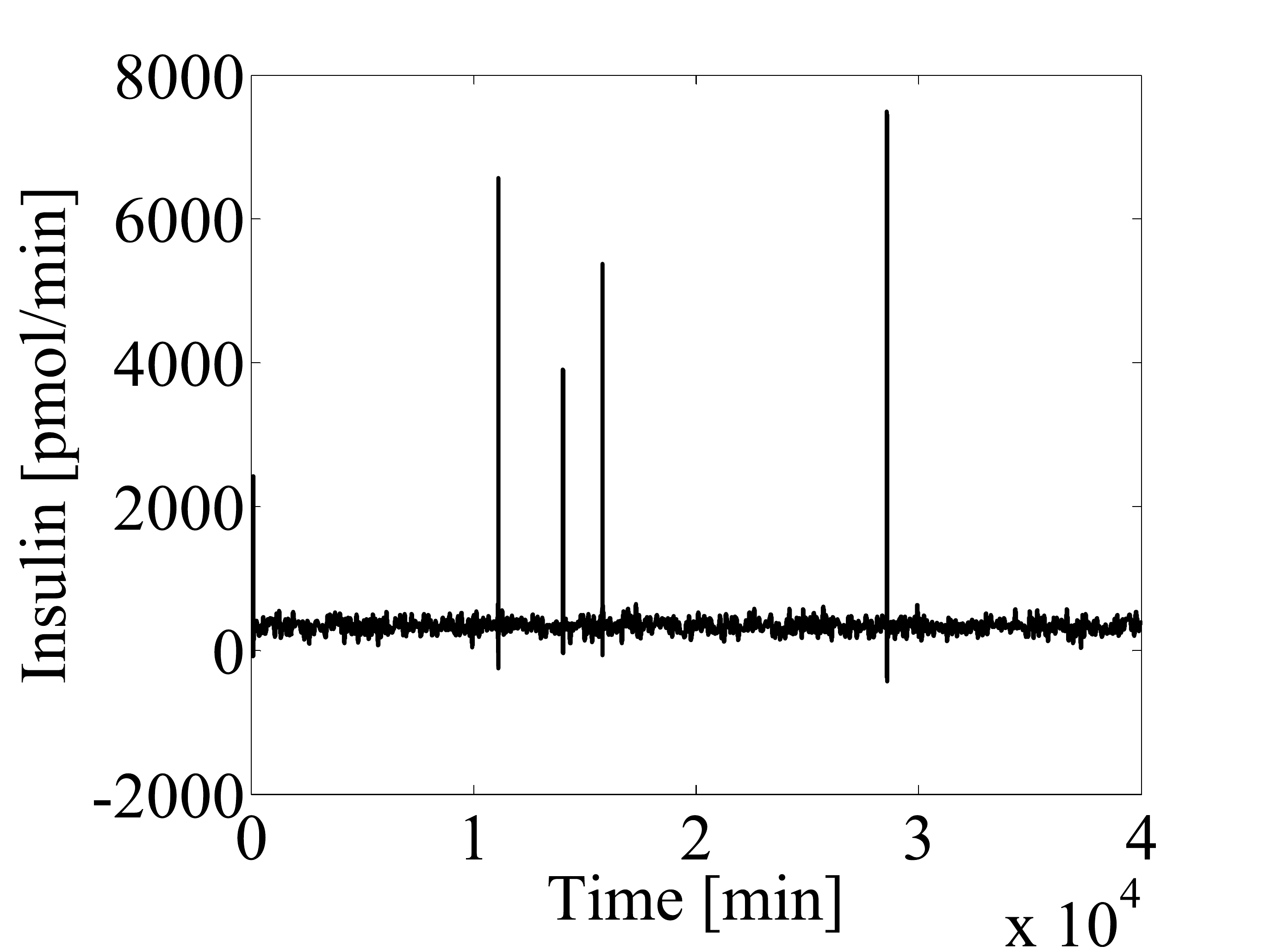}
\caption{Excitation signal: band-limited pulses combined with a random phase multisine signal, at one operating point of the system.}
\label{figexcpulses}
\end{figure}

To reproduce a realistic scenario, the input signal in Figure~\ref{figexcpulses} is used as excitation. It consists of a random phase multisine superimposed on a band-limited pulse signal. The multisine component of the signal accounts for the small variations of the insulin input, while the pulses represent the insulin bolus injected at specific moments (e.g. meal-time, corrections for high glucose levels, and so on). This method of creating a band-limited pulse signal and combining it with a multisine is a procedure that enhances a realistic signal for the purpose of system identification \cite{Dha11}. 

This signal is again applied at 12 different operating points. Two periods of the signal are available, each made of 40000 points. In this case, the sampling time is equal to 1 minute.

The results obtained with the new dataset are summarized in Table~\ref{tabrespulses}.

\begin{table}[!t]
\begin{center}
\caption{\label{tabrespulses} {Results of the nonlinear models: band-limited pulses and multisine excitation. The average relative error over the 12 operating points on the validation set is reported, together with the minimum and maximum relative error values (in brackets) and the number of model parameters.  As a comparison, also the results of a fourth order BLA are shown.}}
\begin{tabular}{ c  c  c }
\hline
\textbf{Model} & \textbf{Average $e_{\text{rel}}$ (min -- max) $\%$} & \textbf{$\#$ parameters} \\
\hline
Wiener & 14.3 (6.5 -- 23.6) & 15 \\
\hline
Wiener-NN & 12.8 (4.2 -- 27.4) & 22 \\
\hline
Wiener-Schetzen & 11.9 (5.8 -- 22.3) & 135 \\
\hline
\hline
BLA & 54.1 (13.7 -- 105.7) & 9 \\
\hline
\end{tabular}
\end{center}
\end{table}

Due to the very large amount of data in the new dataset, PNLSS and NN-based NLSS models could not be obtained, since for these methods the required computational time becomes prohibitive. The reason for this is the very high computational complexity required for the nonlinear optimization step for both methods. As already mentioned in Sections \ref{secPNLSS} and \ref{secNN-NLSS}, the calculation of the Jacobian is carried out in a recursive way at each iteration of the Levenberg-Marquardt algorithm, and this results in a drastic increase of the total computational time.

It can be observed that also in this example the nonlinear models yield very good results: the average error values are four times smaller than the error of the BLA.

\begin{figure}[!t]
\centering
\includegraphics[width=12.5cm]{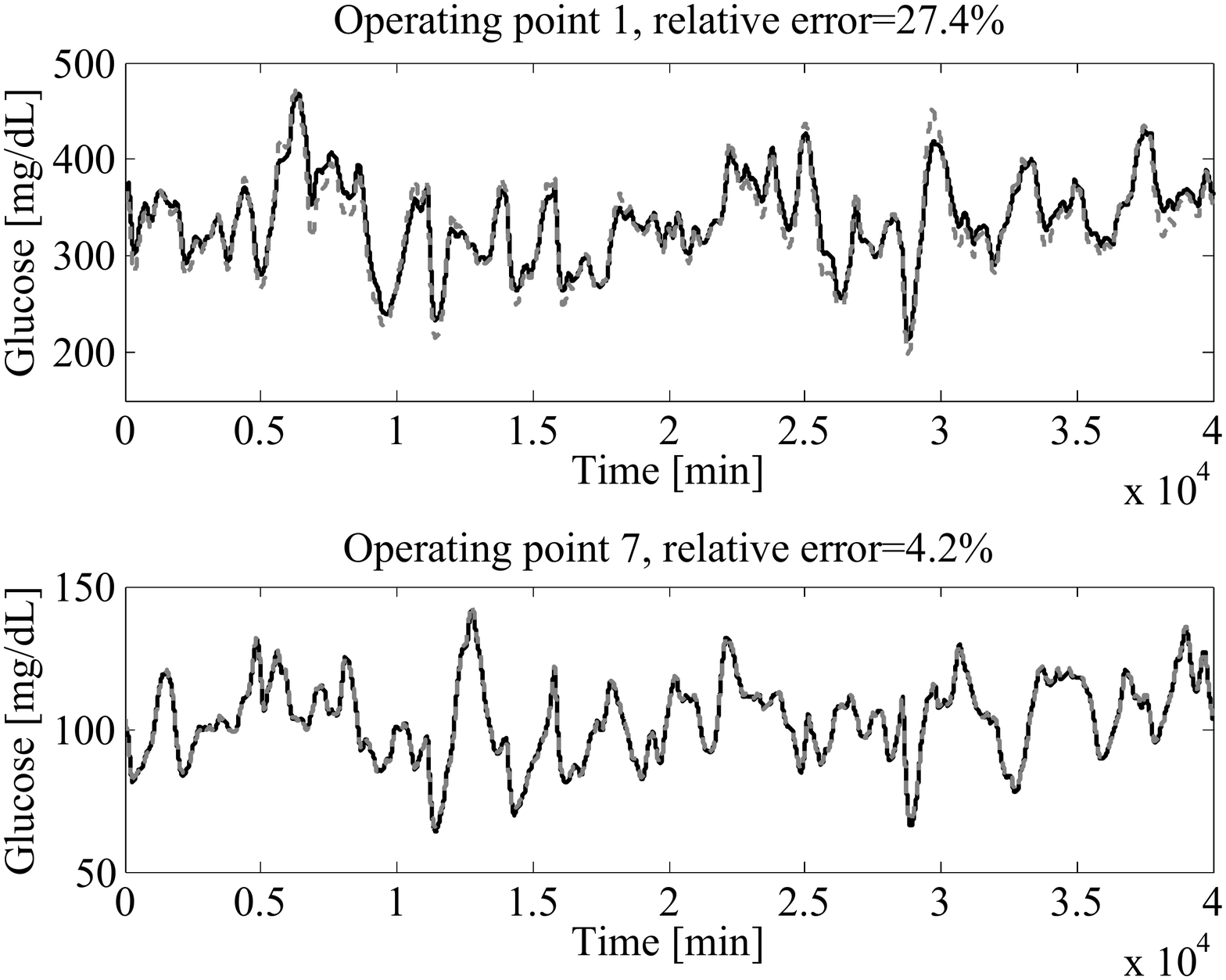}
\caption{Results of the Wiener-NN model: band-limited pulses and multisine excitation. True output (solid black line) and modeled output (dashed gray line) for operating points 1 and 7 are shown.}
\label{modeloutput}
\end{figure}

In Figure~\ref{modeloutput}, the true output and the modeled output for the Wiener-NN are shown for two regions: operating point 1, for which the model performs worst (highest error value: $27.4\%$), and operating point 7, for which the model performs best (lowest error value: $4.2\%$).

The nonlinear model approximates very well the input/output behavior of the system even at operating point 1, and at operating point 7 the modeled output completely overlaps the true output.

\section{Discussion}
\label{sec7}

When comparing the different nonlinear identification methods that have been applied to model the insulin-glucose system, the neural network nonlinearity seems to be more suited to describe the nonlinear behavior of the system than the polynomial function. This can be observed when comparing the Wiener and Wiener-NN models on one hand, and the PNLSS and the NN-based NLSS models on the other hand. Models containing sigmoidal nonlinearities are characterized by lower average error and lower minimum error values.

Note, however, that for the neural network nonlinearity a nonlinear optimization is required, while a simple least squares problem is solved when using polynomials.

Wiener models are simple models that are quite parsimonious in terms of number of parameters, while the NLSS and (especially) the Wiener-Schetzen approaches result in general in more complex models, with a significantly higher number of parameters. 

Wiener-Schetzen models can yield very high accuracy (at specific operating points), although in the multisine excitation case their overall performance is the worst. However, they obtain the best results in terms of average error in the more realistic scenario, with the pulses and multisine excitation signal.

The NLSS methods achieve low error results, but they show severe limitations when the size of the dataset used for estimation becomes very large. Moreover, they seem to be more sensitive to local minima problems.

In summary, when looking at both error values and number of parameters, the Wiener-NN approach seems to offer in both examples the best trade-off between model accuracy and simplicity.

For the implementation of a model-based controller as a fundamental block in the development of the artificial pancreas for T1DM patients, Wiener models seem to represent a good option. The model to be chosen for this specific application should in fact be very accurate to guarantee that no damage is caused to the patient, and simple enough to make the controller implementation easy. 

%

\begin{figure}[!t]
\centering
\subfigure[]{
\includegraphics[width=6cm]{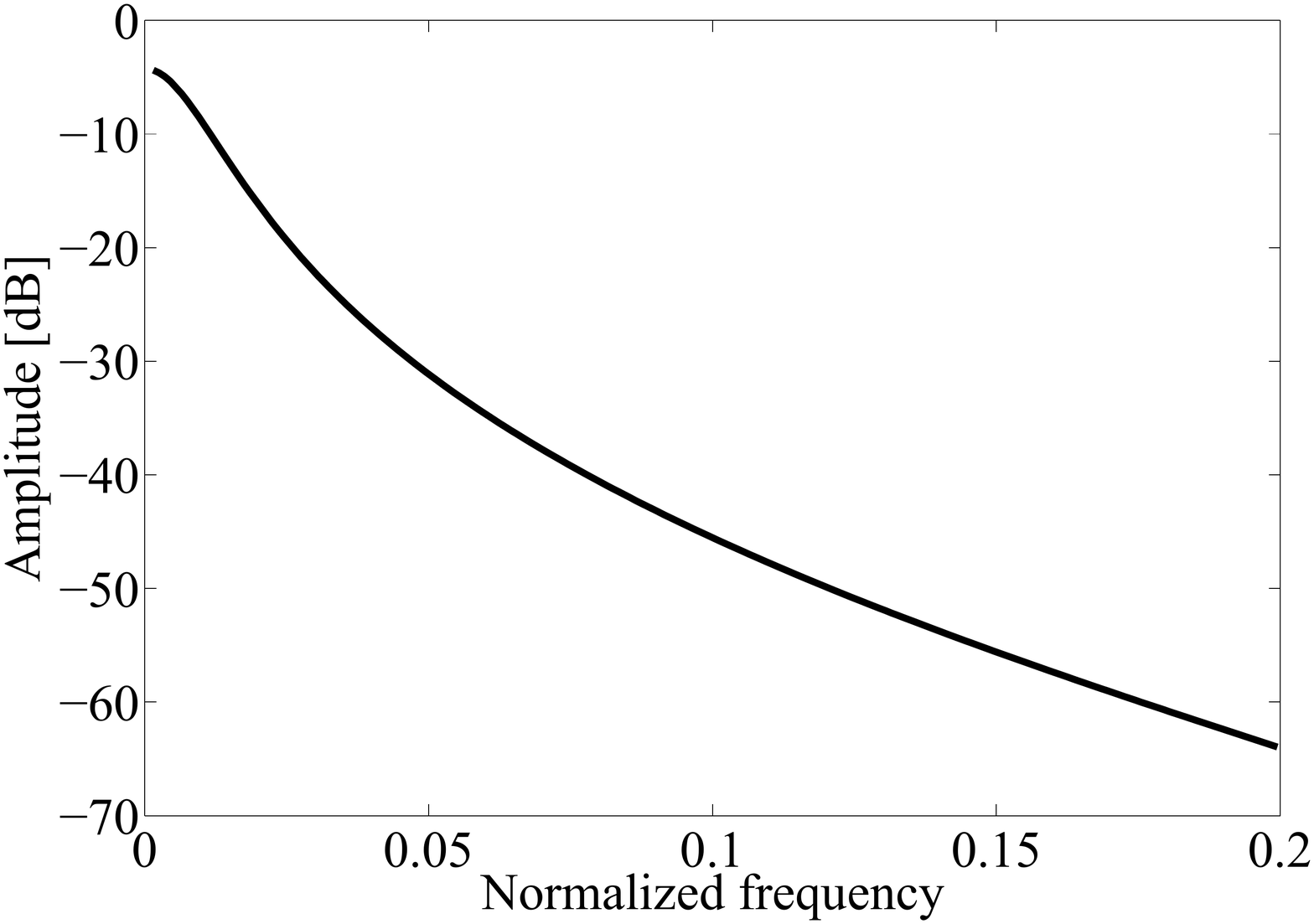}
\label{figlinear}
}
\subfigure[]{
\includegraphics[width=6cm]{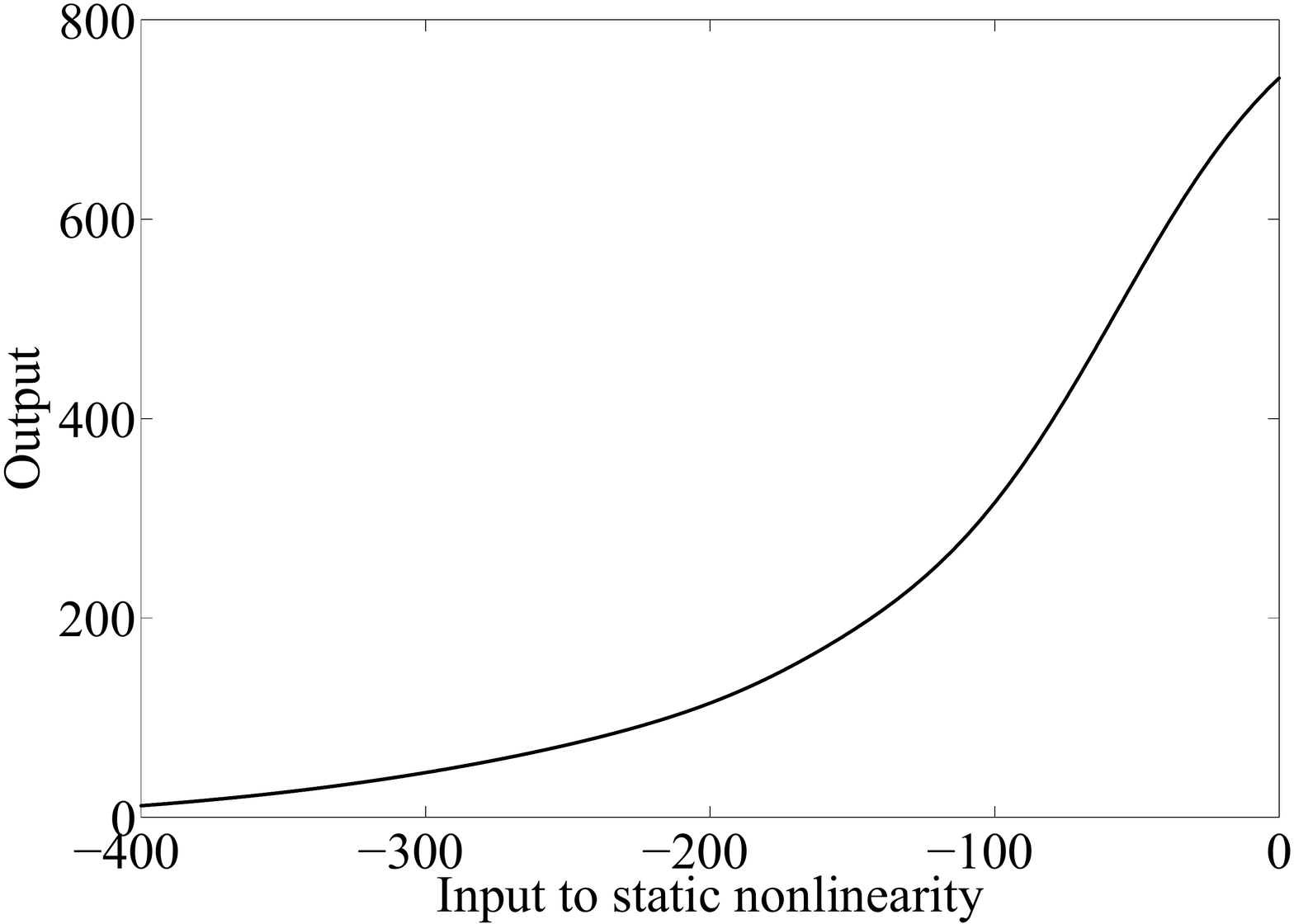}
\label{figstaticNL}
} 
\caption{Wiener-NN model: Magnitude response of the linear block (a) and static nonlinearity (b).}
\label{figlinstatic}
\end{figure}

Finally, the nonlinearity present in the model should be quite easy to invert so that a linear controller can be used, which simplifies the overall control design problem. Both polynomial and sigmoidal functions can therefore be employed in the static nonlinear block in the Wiener model.
As an example, the magnitude response of the linear block and the static nonlinearity that characterize the obtained Wiener-NN model are shown in Figure~\ref{figlinstatic}.

%

\section{Conclusion}
\label{concl}

The objective of this work was the comparison of advanced nonlinear identification methods on a simplified glucoregulatory system modeling example. Several block structures and nonlinear state-space models have been considered to tackle this problem, yielding a significant improvement in terms of accuracy when compared with linear dynamic models. The different methods have been compared on two identification examples, in which the input/output data were generated to simulate the behavior of the insulin-glucose subsystem. As a next step, the obtained nonlinear descriptions can be employed to implement a model-based controller as a first step in the development of the artificial pancreas for diabetes patients.

\section*{Acknowledgements}

This work is sponsored by the Fund for Scientific Research (FWOVlaanderen), the Flemish Government (Methusalem Fund, METH1), the Belgian Federal Government (IAP VI/4), the ERC Advanced Grant SNLSID, and by the University of Girona through the BR-UdG research grant to A. Abu-Rmileh.

\bibliographystyle{IEEEtran}
\bibliography{CTA-2014-0534bibtex}







\end{document}